\documentclass[12pt,preprint]{aastex}
    \usepackage{graphicx,epsfig}
    \usepackage{lscape}
    \usepackage{rotating}
\newcommand{\bq}{\begin{equation}}
\newcommand{\eq}{\end{equation}}

\def\gtsim{\lower.5ex\hbox{$\buildrel > \over\sim$}}
\def\ltsim{\lower.5ex\hbox{$\buildrel < \over\sim$}}

\def\apjl{ApJL}
\def\apj{ApJ}

\def\mnras{MNRAS}

\def\aap{A\&A}

\def\nat{Nature}

\shorttitle{Modeling the light curve of SCP06F6}
\shortauthors{Chatzopoulos, Wheeler,Vinko}
\begin{document}
\title
{Modeling the light curve of the transient SCP06F6}
\author {Emmanouil Chatzopoulos\altaffilmark{1}, J. Craig Wheeler
\altaffilmark{1}, J. Vinko\altaffilmark{1,2}}
%%%  author names
\authoremail{manolis@astro.as.utexas.edu, wheel@astro.as.utexas.edu, vinko@astro.as.utexas.edu}
\altaffiltext{1}{Department of Astronomy, University of Texas at
Austin, Austin, TX, USA}
\altaffiltext{2}{Department of Optics and Quantum Electronics, University of Szeged,
Hungary}

\begin{abstract}
We consider simple models based on core collapse or pair-formation supernovae
to account for the light curve of
the transient SCP06F6. A radioactive decay diffusion
model provides estimates of the mass of the required radioactive nickel
and the ejecta as functions of the unknown redshift. An opacity change such as
by dust formation or a recombination front may account for the rapid
decline from maximum. Within this
class of model, the redshift must be less than $z\sim 1$ or the
nickel mass would exceed the total mass of the ejecta, the 
radiated energy would exceed the kinetic energy and kinematic 
and photometric estimates of the radius would disagree. We 
particularly investigate two specific redshifts: $z=0.143$, for 
which Gaensicke et al. (2008) have proposed that the unidentified 
broad absorption features in the spectrum of SCP06F6 are 
C$_{2}$ Swan bands, and $z=0.57$ based on a crude agreement with 
the Ca H\&K and UV iron-peak absorption features that are characteristic
of supernovae of various types. For the lower redshift, we obtain 
a nickel mass of 0.3~M$_{\odot}$ and an ejected envelope mass of 
$\sim$ 38~M$_{\odot}$, while for the latter case we find 
4.8~M$_{\odot}$ and 20~M$_{\odot}$, respectively, for fiducial
parameters.  The kinetic 
energy of the ejecta, while dependent on uncertain parameters,
is generally large, $\sim 10^{52}$ erg, throughout this range of
redshift. {\bf The ejected masses and kinetic energies are smaller for a 
more tightly constrained model invoking envelope recombination.}
We also discuss the possibilities of circumstellar matter (CSM) shell diffusion 
and shock interaction models. In general, optically-thick CSM diffusion 
models can fit the data with the underlying energy coming from an
energetic buried supernova. Models in which the CSM is of lower 
density so that the shock energy is both rapidly thermalized and 
radiated tend not to be self-consistent. 
%The exceptionally high X-ray luminosity of SCP06F6 can be interpreted
%in the framework of a lower density shocked CSM that lies beyond
%the optically thick CSM component. 
We suggest that a model of SCP06F6 worth futher exploration is one 
in which the redshift is $\sim$ 0.57, the spectral features are Ca 
and iron peak elements, and the light curve is powered by the 
diffusive release of a substantial amount of energy from nickel decay
or from an energetic supernova buried in the ejecta of an LBV-like event.

\end {abstract}


% sc1
% SECTION 1
\vskip 0.57 in
\section{INTRODUCTION}\label{intro}

Recently discovered very luminous supernovae, such as SN~2006gy
(Quimby 2006; Smith et al. 2007), SN~2005ap (Quimby et al. 2007a),
SN~2006tf (Quimby, Castro \& Mondol 2007; Quimby et al. 2007b; 
Smith et al. 2008) and SN~2008es
(Yuan et al. 2008; Gezari et al. 2008; Miller et al. 2008)
introduce evidence for new supernovae phenomena. The 
discovery of SN~2006gy triggered a rich discussion
of the nature of this event and a number of models were
proposed to interpret the observed light curve (Ofek et al. 2007;
Smith et al. 2007; Smith \& McCray 2007; Woosley, Blinnikov \& Heger 2007; 
Angoletto et al. 2008).  SN~2006gy is classified as a Type IIn event that peaked at
V$\sim$22 70 days after the explosion and radiated away more
than 10$^{51}$ergs of energy. Radioactive $^{56}$Co decay
fits of the light curve of SN~2006gy imply extraordinary
amounts of initial nickel mass of the order of 22~M$_{\odot}$
(Smith et al. 2007). The detection of soft unabsorbed X-rays
by $XMM-Newton$ indicated that the extended Circumstellar Matter (CSM) environment of SN~2006gy
is of low density. Angoletto et al. (2008) challenged this
hypothesis by considering highly opaque clumps distributed around
SN~2006gy from which they derive a nickel mass estimate of
about 3~M$_{\odot}$ (see Smith et al. 2009 for an extensive summary). 
SN~2006gy might be consistent with a model
of a pair formation supernova (Rakavy \& Shaviv 1966; Barkat, Rakavy \& Sack 1967;
Smith et al. 2007), but SN~2005ap, which is even brighter
but with a narrower light curve, cannot be (Quimby et al. 2007a).

Recently, another apparently ultraluminous transient event
was presented by the Supernovae Cosmology Project,
SCP06F6 (Barbary et al. 2008). The possible high brightness,
slow rise to maximum (100 days), and strange spectral features
of SCP06F6 are still under debate. There is no detected host
galaxy consistent with the position of SCP06F6 although examples
of low mass star-forming galaxies have been found that might
be consistent with the upper limits (Dolphin et al. 2001).
%Additionally, there were two $XMM-Newton$ observations of SCP06F6,
%before and after maximum, that show an X-ray flux two orders of
%magnitude larger than that from normal core-collapse SNe
%(Gaensicke et al. 2008). 
Thus SCP06F6 could define a new class
if it is actually associated with a supernova explosion.
Due to the peculiar spectral appearance of SCP06F6 and the
lack of an identified host galaxy, its redshift remains uncertain.

We discuss the uncertainties in the redshift, spectrum and luminosity
in \S 2. In the present work we examine a range of possibilities for the redshift and two
special cases. We present a basic radioactive decay diffusion model plus the possible effects
of a recombination front and dust formation in \S 3 and use it
to place constraints on the distance, ejected mass, nickel mass,
and kinetic energy in \S 4. In \S 5, we present models based on interaction
of a supernova shock with circumstellar matter. A discussion 
is presented in \S 6.

\section{UNCERTAIN REDSHIFT, SPECTRUM AND LUMINOSITY}\label{redspec}

Barbary et al. (2008) discussed the
possibility of the extragalactic origin of SCP06F6 since
it has a small projected distance from the center of the
galaxy cluster CL 1432.5+3332.8 which has a redshift of 1.112.
They noted that at this redshift, the absorption feature at
5890~\AA\ aligns with MgII~$\lambda \lambda$ 2796, 2803,
but were unable to identify the other features. In particular, there are no
features readily identifiable with hydrogen. In this work
we argue that the redshift cannot be above $z\sim 1$
through various constraints on the ejecta mass, luminosity
and radius. 

One special case corresponds to $z=0.143$, as
proposed by Gaensike et al. (2008) by considering SCP06F6
as the explosion of a massive carbon-rich star with a cool
and optically thick atmosphere. Gaensicke et al. (2008)
propose that the broad ``absorption" features present
in the blue end of the spectrum of SCP06F6 (Figure 1) are
molecular C$_{2}$ bands (the Swan bands) originating in the
cool, optically-thick atmosphere of a carbon-rich progenitor.
The presence of carbon molecules constrains the temperature
of the atmosphere of the progenitor to be cool so that they
are not destroyed. Gaensicke et al. (2008) find that carbon
star spectra redshifted by 0.143 provide a decent fit to
the combined VLT+Keck spectrum of SCP06F6 (see also Soker,
Frankowski \& Kashi 2008). {\bf Other
features present in carbon star spectra, however, such as CN ($\lambda$7900 \AA\ rest frame, 
Downes et al. 2004),
do not seem to appear in the smoothed co-added spectra of SCP06F6 
(VLT+Subaru+Keck). Other CN bands would fall too far to the blue
to be constrained by the available spectral data.} We return to this issue in \S 6.

We also investigate the case for
$z\sim 0.57$ where we can crudely identify three of the four 
broad absorption features by redshifting the spectrum of the
local Type Ia SN~1992A or a template representing a Type IIP
(Gililland et al. 1999) as seen in Figure 1. A redshift of 
$z\sim 0.57$ aligns the Ca II H\&K absorption with the minimum 
at $\sim$ 5900 \AA\ in SCP06F6. The two bluest minima crudely 
correspond to minima in the UV spectra of SN~1992A and the SN IIP
template caused by overlapping lines of iron-peak elements 
(Harkness \& Wheeler 1987; Kirshner et al. 1993). Similar features 
are seen in other supernovae of various spectral types with UV 
spectra (Bufano et al. 2009). The agreement is not good with the minimum at about
5300 \AA\ in SCP06F6 where the SN II template has a peak. Note 
that at this redshift the continuum slopes redward of 4000~\AA\ (but not the spectral features)
also roughly agree. 
The comparison in Figure 1
does not suggest that SCP06F6 is either a Type Ia or a Type II, but 
that SCP06F6 might have Ca II and iron-peak absorption in the 
rest-frame UV moving with velocities typical of supernovae.
For reference, we also indicate in Figure 1 where H$\alpha$ and H$\beta$
would fall for this redshift. H$\alpha$ would fall off the red end of the spectra
and H$\beta$ would fall near the gap in the spectra that is contaminated by
telluric lines. For this redshift the lack of evidence for H in the spectrum cannot
be taken as firm evidence for lack of hydrogen in the ejecta. The same is not true
for the redshift of 0.143 where hydrogen might be present in emission or absorption
for a variety of models, including those not based on supernovae.

To approximate the bolometric lightcurve of SCP06F6 we use the VLT spectrum
at maximum (Barbary et al. 2008) scaled to the i- and z- band fluxes observed at
the same epoch. We then approximate the SED of the object with a triangle
peaked at 6000 \AA\, that extends from 3000 to 12,000 \AA\ .
We calculate the integral to get the quasi-bolometric
flux at maximum light, $F_{bol}$. This approximate
method gives $F_{bol,max}=3.09 \times 10^{-14}$ $erg~s^{-1}cm^{-2}$ which is
in good agreement with the estimated value of $2.5 \times 10^{-14}$ $erg~s^{-1}cm^{-2}$
given in Barbary et al. (2008).
We then scale the flux to the other epochs
using the photometry in Table 1 of Barbary et al. (2008) assuming that the shape of the
SED is roughly constant.
We use the following
formula to obtain bolometric luminosities:
\begin{equation}
L_{bol}=F_{bol} 4 \pi r^{2},
\end{equation}
where $r$ is the luminosity distance of the object which is equal
to $r=(1+z/2) \times cz/H_{0}$, where $c$ is the speed of light, $z$ is the redshift
of the object and $H_{0}=70$ $km~s^{-1}Mpc^{-1}$ is the Hubble constant. This expression
for $r$ corresponds to an empty universe and
is a good approximation to the full $\Lambda$-CDM expression.

%Finally, we explore arguments based on circumstellar (CSM)
%interaction models. 

%It should be noted that we only have SDSS i and z band 
%measurements of the SCP06F6 light curve, and we thus have 
%only lower limits for the bolometric luminosity of SCP06F6.
% In addition, we do not have observations in later parts of
%the light curve that would give an additional estimate of
%the initial nickel mass
%using the Sutherland \& Wheeler (1984) expression for the Co decay tail.

%s2
% SECTION 2
\section{A SIMPLE DIFFUSION MODEL FOR SCP06F6}\label{diffmod}

\subsection{{\it Radioactive diffusion model}}\label{radif}

We adopt the Arnett (1982) radioactive decay diffusion model
as generalized by Valenti et al. (2008; see also Soderberg et al.
2008) to allow for both nickel and subsequent cobalt decay. We also allow for
gamma-ray leakage.
This simple model assumes a diffusive medium with a uniform
density profile and a photosphere that expands linearly in time. 
In general, the solutions we find for SCP06F6 imply large nickel and ejecta masses. 
With these assumptions, we adopt the
following expression for the diffusive luminosity released as
a result of radioactive decay:
\begin{equation}
L(t)=M_{Ni}e^{-x^{2}}[(\epsilon_{Ni}-\epsilon_{Co}) \int_0^x2ze^{z^{2}-2zy}dz
  + \epsilon_{Co}\int_0^x2ze^{z^{2}-2yz+2zs}dz](1-e^{-At^{-2}}),
\end{equation}
where $x=t/t_{m}$, $t_{m}$ is {\bf the rise time to maximum (Arnett 2008)},
$y=t_{m}/2t_{Ni}$ with $t_{Ni}=$8.8 days,
$s=t_{m}(t_{Co}-t_{Ni})/2t_{Co}t_{Ni}$ with
$t_{Co}=$111.3 days, $M_{Ni}$ is the initial nickel mass, 
$\epsilon_{Ni}=3.9 \times 10^{10}$ $erg~s^{-1}g^{-1}$
and $\epsilon_{Co}=6.8 \times 10^{9}$ $erg~s^{-1}g^{-1}$ are the energy generation rates
due to Ni and Co decay. The factor
$(1-e^{-At^{-2}})$ accounts for the gamma-ray leakage, where large $A$
means that practically all gamma rays are trapped. The gamma-ray optical depth of 
the ejecta is taken to be $\tau_{\gamma}=\kappa_{\gamma} \rho R=At^{-2}$, assuming spherical uniform density
ejecta with radius $R=vt$ and the Ni/Co confined in the center. 
This yields $A=(3\kappa_{\gamma} M_{ej})/(4 \pi v^{2})$ which is controlled by the
gamma-ray opacity $\kappa_{\gamma}$. The $t^{-2}$ scaling follows from
homologous expansion which is one of the basic assumptions of the simple analytic
models that we adopt here. In general we find the effect of gamma-ray leakage to be small
for the large masses implied in our models.

Thus we have three main fitting parameters that determine the
nature of the model light curve: the initial $^{56}$Ni mass
that determines its peak, the rise time to maximum $t_{m}$
that determines its width and $A$ that determines the amount of gamma
ray trapping and corresponds to a gamma-ray opacity.
It is characteristic that, within
this class of model, the nickel mass comes in only as a
scaling parameter for the amplitude of the luminosity output.
The more distant the source is, the brighter it is, since
its intrinsic (rest frame) luminosity is higher 
compared to the observed luminosity.
We thus expect a higher initial $^{56}$Ni mass at higher redshifts.

Unlike the maximum brightness, the width of the light curve in the object
rest frame decreases with respect to its observed value, since the rest frame rise
time, $t_{m,rf}$, decreases with redshift:
\begin{equation}
t_{m}=\frac{t_{m,ob}}{1+z},
\end{equation}
where $t_{m,ob}$ is the observed {\bf rise} time.
Assuming that the photosphere expands with approximately
constant speed, $v_{ph}$, we adopt the ejecta mass as a function
of {\bf the observed rise time} given by Valenti et al. (2008):
\begin{equation}
M_{ej}(z)=\frac{3}{10} \frac{\beta c}{\kappa} v_{ph} \frac{t_{m,ob}^{2}}{(1+z)^{2}},
\end{equation}
where $\beta$ is an integration constant equal to about 13.8 (Arnett 1982;
Valenti et al. 2008), $\kappa$ is the mean opacity, and we have used
$v_{ph} = (10/3 E_{KE}/M_{ej})^{1/2}$, as appropriate for the
outer edge of a homologously expanding sphere of constant density
with ejecta mass $M_{ej}$ and kinetic energy $E_{KE}$. The ejecta
mass is set by the rise time to maximum that depends on the redshift of the
object. Thus we obtain a scaling of the ejecta mass with
redshift within the context of this model for SCP06F6. The
photospheric velocity of supernova ejecta is estimated
from the width of lines in their spectra. For Type II supernovae it
typically varies within the range $4000-6000$~$km~s^{-1}$, and for
Type Ia supernovae it is typically $10,000 - 15,000$~$km~s^{-1}$.
We note that the features of SCP06F6 are generally broader than those
of the Type IIP template of Figure 1 and perhaps more represented by 
the features of the Type Ia. Barbary et al. (2008) give
$v_{ph} \simeq 12,000$~$km~s^{-1}$. We adopt $v_{ph}=10,000$~$km~s^{-1}$ as a 
fiducial value and assume this to refer to the photospheric velocity 
in the rest frame of the supernova {\bf for the models presented in
\S 4.} With assumed values of opacity we
attempt to fit the observed light curve and to obtain constraints
on physical properties of SCP06F6 that will in turn scale as
a function of the redshift.

\subsection{{\it The effects of recombination}}\label{recomb}

The light curve presented by Barbary et al (2008) is
roughly symmetric about the peak. The fact that SCP06F6 shows a symmetric light 
curve is not in contradiction with observed
supernovae characteristics. There are cases such as the Type Ib 
supernova SN2005bf (Tominaga et al. 2005) for which the second peak has a very symmetric light curve.
As we will show below, our simple radioactive diffusion models
have difficulty accounting for the low values of the points
100 days after peak and hence reproducing the symmetric light curve.
There are several factors that might account for this discrepancy.
The fundamental assumption of the model of \S 3.1 is that the 
ejecta are optically thick. That may not be true at the last observed epoch. 

An expression for the
optical depth at maximum light can be written as:
\begin{equation}
\tau_{max} \sim \kappa \bar\rho R \sim \frac{3 \kappa M_{ej}}{4 \pi R^{2}}.
\end{equation}
Using the expression for $\kappa M_{ej}$ from Equation 4 and
the kinematic radius $R_{k,max} = v_{ph} t_{max}/(1 + z)$ gives
$\tau_{max} \simeq c/v_{ph}$, 
independent of redshift. For a photospheric velocity of 10,000 km s$^{-1}$,
the optical depth would be about 30, thus justifying the
assumption of large optical depth at maximum light. 
Using the black body radius gives a similar estimate of large
optical depth, but one that varies with redshift.   

As the ejecta expand the optical depth decreases as $\tau \simeq t^{-2}$. For
constant opacity
the optical depth at epoch 11 of Barbary et al. (2008) (Table 1) is thus
$\tau_{11} \simeq \tau_{max} (t_{max}/t_{11})^{2} \simeq 30/3 \simeq 10$ for
$\tau_{max} \simeq 30$, $t_{max}=$~100 d and $t_{11}=$~170 d.
If the opacity dropped by a factor of several after maximum due to recombination
in some element or elements, the light curve might decline more rapidly than
our models. 

To incorporate the effects of recombination we use the model developed by Arnett \& Fu (1989)
to fit the light curve of SN1987A. This model considers the effects of the
recombination of hydrogen or helium. After peak luminosity
the recombination front recedes inwards and the optical opacity drops dramatically. Thus
the thermal energy generated by the presumed shock diffuses out more quickly and the decline is faster than it would
be without recombination. 
{\bf The light curve that results from this model depends on the following input parameters:
the initial radius of the progenitor $R_{0}$, the nickel mass yield $M_{Ni}$, the mass of the ejecta
$M_{ej}$, the kinetic and the thermal energy $E_{kin}$ and $E_{th}$ respectively, the ionization temperature
$T_{ion}$ and the optical opacity $\kappa$. The values that we used to obtain the fits shown in
Figure 2 are summarized in Table 2. Also given in Table 2 is the mean ejecta
velocity, $v_{mean}=\sqrt{2E_{kin}/M_{ej}}$ implied by the 
model fits. The amount of nickel needed to power the event in the recombination
model is less than it is if the recombination effects are not taken into account, as expected. Specifically
$M_{Ni,rec}/M_{Ni}\simeq0.6$ for both redshift models, where $M_{Ni,rec}$ is the mass of nickel 
that is required to power the peak of the light curve if the recombination effects are taken into account.}

As seen in Figure 2, {\bf a recombination model} can produce a roughly symmetric light
curve around the peak that subsequently follows the cobalt decay.
{\bf In our model for SCP06F6 we consider the composition of the ejecta unknown and we adopt the Thompson scattering
opacity values of 0.4 $cm^{2}~g^{-1}$ at $z=$0.143, (a typical value for a H-rich envelope) and 0.2 $cm^{2}~g^{-1}$
at $z=$0.57 (a typical value for a H-poor envelope).}
As noted in \S 2 there is no evidence for hydrogen
or helium in the spectrum of SCP06F6, but if it is at a redshift of 0.57, then
the H$\alpha$ line would be redshifted out of the range of the spectral converage by VLT
and Subaru as shown in Barbary et al. (2008) and H$\beta$ would be contaminated by telluric lines (Figure 1).
{\bf We are not able to accurately reproduce the light curve of SCP06F6, even when the recombination effects
are taken into account, especially due to the fact that we cannot fit Epoch 11 of Barbary et al. (2008).
It should be noted that the purpose of the recombination models that we consider is to illustrate
the effects of this process on the light curve of a SN, and to show that recombination does tend to make
the light curve more symmetric around the peak. More detailed models accounting for the density, composition and
optical depth structure and appropriate radiative transfer are beyond the scope of this work.}

\subsection{{\it The effect of dust}}\label{dusteff}

There could also be dust formation at late epochs. At Epoch 11
(Table 1, Barbary et al 2008) the observed flux is $\sim$ 7 times lower
than the flux that is predicted by the model {\bf (insensitive to redshift).} Thus we can estimate
the optical depth due to a hypothetical dust absorption using the formula $F_{11}=F_{0,11}e^{-\tau_{dust}}$, where
$F_{11}$ is the flux at Epoch 11, $F_{0,11}$ is the flux at epoch 11 as predicted by the radioactive
decay diffusion model and $\tau_{dust}$ is the dust optical depth.
This yields $\tau_{dust}\simeq 1.9$ if we consider a spherical, optically thin dust shell at the
outer edge of the SN ejecta with radius $\simeq$ $R_{11}$ $=v_{ph} t_{11}$ and thickness $dR << R_{11}$.
The required dust mass is then $M_{dust,11}=4 \pi R_{11}^{2} \rho_{dust} dR$, and the dust optical depth
$\tau_{dust}=\kappa_{dust} \rho_{dust} dR$. Combining
these equations yields $M_{dust}=\tau/(\kappa_{dust} 4 \pi R_{11}^{2})$ where $\kappa_{dust}$
is the opacity of the dust grains. We adopt the value $\kappa_{dust}=10^{4}$ $cm^{2}~g^{-1}$ 
as a gross estimate, using the Draine (2003) model extinction curves for 
Milky way dust grain composition. For this choice of parameters
we estimate the dust mass to be $\sim 3 \times 10^{-4}$~$M_{\odot}$. This value is in agreement
with values that are estimated for other Type-II supernovae (Kotak et al. 2005; Kotak et al. 2009).

If the dust formation took place well inside the ejecta, for example at $v \simeq$~1000 km~s$^{-1}$, as in Type II-P 
SNe (Kotak et al. 2009), then the dust mass would be 100 times higher than the value estimated above, on the 
order of 0.02$M_{\odot}$. This is somewhat higher than has been observed in other SNe, but still not unrealistic, 
taking into account that the dust masses derived in this way are only lower limits. If the dust distribution is 
not uniform and optically thin but rather it is distributed in optically thick clumps (Ercolano et al., 2007), 
then the total dust mass could be much higher than that derived from the optically-thin model.

Note that the energy absorbed by dust is expected to be re-radiated as mid-IR photons after the dust particles 
(or clumps) are heated up to $T_{dust} \sim 600$ -- $1000$ K. Thus, strictly speaking, the true bolometric 
luminosity remains unchanged in this case. In \S 2, we estimated the bolometric luminosity by assuming that 
all radiation comes from the optical and near-IR regime. The Arnett model (Equation 2) also uses essentially  
the same assumption due to the thermalization of the gamma-rays. Within this context the optical photons 
absorbed by dust can be considered as ``lost," even though their energy is expected to re-appear as a mid-IR excess, 
outside the original SED.

\section{CONSTRAINTS ON SCP06F6}\label{diffmod}

%s2.1
\subsection{{\it General redshift constraints}}\label{general}

{\bf Here we calculate the redshift dependence of the physical parameters of the Arnett model
given in Equation 2, without recombination.}
Figure 3 shows how the ejecta mass and the nickel mass depend on redshift
based on fits to the light curve for various choices of
the mean opacity. {\bf The recombination models have the same redshift
dependence, but the nickel masses are about 40\% lower, the ejecta masses are about 30\% lower and the kinetic
energies are lower by a factor of 2 to 3 compared to our fiducial models with the same opacity.}
The ejecta mass {\bf for the basic radioactive diffusion models} is calculated using Equation 4 adopting
$t_{m,ob}=100$~d for the rise time to maximum in the observer's frame {and adopting $v_{ph}=10,000$ km s$^{-1}$
as the fiducial velocity.}
The nickel mass is calculated by assuming $t_{max}=t_{m}$ 
and solving Equation 2 for $M_{Ni}$ given the observed value
of maximum luminosity, which scales with redshift according to Equation 2.
We consider values of the mean opacity ranging
from 0.05 $cm^{2}~g^{-1}$ as might be representative of a metal-rich
composition (Sutherland \& Wheeler 1984; Soderberg et al. 2008) up to
0.40 $cm^{2}~g^{-1}$ as might be representative of electron scattering
in a pure hydrogen plasma. 
Figure 3 also shows the best fitting results for the nickel mass
and the ejecta mass in the case of $\kappa$=0.1 $cm^{2}~g^{-1}$ for the same
redshift range and with redshift increments of 0.1 (filled circles
and squares respectively). To determine
the best fitting parameters ($M_{Ni}$, $t_{m}$ and $A$) for each
redshift we developed a simple Monte Carlo chi-square minimization code
that scans through all the parameter space and finds the minimum
$\chi^{2}$ value and the parameters that correspond to that value. As
can be seen in Figure 3, the best
fitting results are in very good agreement with the analytic ones assuming
$t_{m,ob}=100$~d {\bf as a fixed parameter}.

The region above redshift of
$\sim$1.1 is forbidden since the required nickel mass 
would exceed the required ejecta mass to account for
the width of the light curve even for rather small opacity. The
region to the right of the dashed vertical line at $z\sim$0.8 (for 
$\kappa$=0.1 cm$^{-2}$ g$^{-1}$)
is also forbidden in practice because it is unlikely that the nickel
mass exceeds half the total ejecta mass in an astrophysically
realistic model of, for example, a pair-formation or core-collapse
supernova. These redshift limits get tighter for larger mean
opacities or lower photospheric velocities. For electron scattering 
in a pure ionized hydrogen plasma, the redshift would have to be less 
than about 0.65 for this sort of model to be self-consistent. {\bf Invoking recombination
does not change this constraint substantially since $M_{ej}/M_{Ni}$ is about
the same (Tables 1,2).}
A nickel mass of 1M$_{\odot}$ that would be characteristic of a Type Ia 
is obtained at a redshift of about 0.3. At this redshift, the ejecta 
would exceed 30~M$_{\odot}$ for a mean opacity of $\kappa~\ltsim~0.1$ 
$cm^{2}~g^{-1}$, so the explosion is certainly not that of an exploding 
white dwarf.

% new JCW text
Figure 4
shows the scaling with redshift of the kinetic energy, E$_{KE}=1/2 M_{ej} v_{mean}^{2}$,
and the total radiated energy, E$_{rad}$. The kinetic energy is
determined from the ejecta mass, assuming
a mean velocity of $v_{mean}=\sqrt{3/5} v_{ph} = 7,800 $ km s$^{-1}$
and the same range of mean opacity as before. Note that the kinetic energy is
especially sensitive to the assumed photospheric velocity, scaling
as E$_{KE}$ $\propto v_{ph}^{3}$ (from Equation 4). {\bf The more tightly-constrained
recombination models require a lower $E_{kin}/M_{ej}$ and hence
$v_{mean}$ than our fiducial radioactive diffusion models (Table 2). This raises 
the possibility that our fiducial models presented here and in Figure 4 overestimate
the ejecta velocity and kinetic energy.} The radiated 
energy is obtained by integrating the bolometric luminosity 
over the whole light curve. Since a strong
shock distributes energy equally between kinetic energy and
thermal energy in its wake and some thermal energy is likely
to be subsequently converted to kinetic energy by adiabatic expansion
and PdV work, it is unlikely that the radiated energy can exceed the kinetic
energy.  This constraint gives another limit on the redshift,
z $\ltsim$ 1.1 for $\kappa$=0.1 cm$^{-2}$ g$^{-1}$, as shown in
Figure 3.  Note that if the opacity were larger or the photospheric
velocity smaller, this constraint also gets tighter.

%{DISCUSS]
%Gaensicke et al. (2008) give a 0.5-10 keV X-ray flux
%near optical maximum of about 10$^{-13}~erg~s^{-1}~cm^{-2}$. 
%The corresponding X-ray luminosity is also given in Figure 3.
%Over the redshift range considered here this is a very large
%X-ray luminosity, $\sim$100 times that of other supernovae.
%We consider the origin of this luminosity in \S \ref{csmssd}.

%JCW text
An estimate of the radius of the photosphere can be obtained
from kinematics ($R =v_{ph} t$) or emissivity (a blackbody
radius). Figure 5 gives the radius estimated at maximum light in these two ways
assuming a constant rest frame photospheric velocity of
10,000 km s$^{-1}$ and a constant rest frame blackbody
temperature of 5000 K. Note that these two
estimates only agree at a single redshift. Each estimate is
uncertain, so we also denote the range over which the
radii agree within a factor of two or a factor of three.
The latter restricts the range of redshift to z $\sim$ 0.2 -- 0.9.   

Another consistency check follows from noticing that the slope
of the continuum varies little over the three epochs for which
spectra were given by Barbary et al. (2008). We have estimated
the effective temperature in our models at these epochs using
the bolometric luminosity and the kinematic radius
from Figure 5. The effective temperature derived this way (that is
formally 5000 K at z$\sim$0.5) varies by $\ltsim$ 300 K over the
spectral epochs, basically consistent with the observations.

%s4.1

%s4.2
\subsection{{\it Special case 1: $z=0.143$}}\label{z=0.143}

We next examine the special case of z = 0.143 as suggested by
Gaensicke et al. (2008) who fit the spectrum of SCP06F6
with redshifted carbon-rich stellar spectra from the SDSS database.
The result of the analysis for z = 0.143 is shown in the top panel of Figure 2.
The best fit gives a value 
of $M_{Ni}$ = 0.28~M$_{\odot}$ and
$M_{ej}$=37.74 ($\frac{\kappa}{0.1~ \rm cm^{-2} \rm g^{-1}})^{-1}$
($\frac{v_{ph}}{10,000~\rm km~ \rm s^{-1}})$~M$_{\odot}$. The best
fitting $A$ parameter which controls the gamma ray leakage implies 
a small gamma ray opacity $\kappa_{\gamma}=5 \times 10^{-4}$ $cm^{2}~g^{-1}$,
whereas a typical gamma-ray opacity might be $\sim$0.03 $cm^{2}~g^{-1}$ (
Colgate, Petschek and Kriese 1980). Gamma-ray leakage is not an important effect
for the large masses derived.
The initial nickel mass is within the range expected for typical SNe
while the ejecta mass is rather large, due to the fact that SCP06F6
shows a significantly slow rise to maximum and a slow decline.
%JCW addition
Note that the ejecta mass varies inversely with the assumed, and
uncertain, opacity (Equation 4). This mass could be less
if the mean opacity were greater than the value, $\kappa = 0.1$
cm$^{-2}$ g$^{-1}$ assumed here for illustration. {\bf It would also be less for smaller $v_{phot}$.}

The value of the ejecta mass implies a total kinetic energy of
$E_{KE}=2.4 \times 10^{52}$ ($\frac{\kappa}{0.1~ \rm cm^{-2}~ \rm g^{-1}})^{-1}$
($\frac{v_{ph}}{10,000~ \rm km~ \rm s^{-1}})^{3}$ erg while a lower limit 
for the radiated energy is $9.1 \times 10^{49}$ erg.
%However we do not expect the true bolometric luminosity of the object
%to be higher two orders of magnitude so we conclude that in this case
%the conservation of energy is violated. 
%JCW addition
While the estimated radiated energy is roughly in accord with ``normal" 
supernovae, the estimated kinetic energy is very large.  For the 
assumed opacity, the implication is that the initial radius was
rather small compared to the radius of the photosphere at maximum light
so that substantial initial thermal energy was lost to adiabatic
expansion. For a larger opacity and a photospheric velocity more in
the range typical of Type II supernovae, the estimated kinetic and
radiated energies would be more nearly equivalent.

The top panel of Figure 2 also shows the results of a model that incorporates the effects
of recombination on the light curve (\S 3.2) for this redshift case. {\bf The recombination
temperature was found to be $T=$~5500~K with $\kappa=$0.4~$cm^{2}~g^{-1}$ which is appropriate for a H-rich
atmosphere, although there is no evidence for the presence of H in the spectrum if this were
the correct redshift. For this model the nickel mass, 
the ejecta mass and the kinetic energy are lower than for the basic model, as mentioned above.} 
The kinetic energy, and hence implicitly the mean
velocity, was varied to produce the dashed curve in the top panel of Figure 2. Given the uncertainties,
no attempt was made to determine the ``best fit" over the full parameter range. This
model shows that recombination could help to produce a roughly symmetric light curve
and that this physics, in principle, is relevant to SCP06F6, {\bf although it was not possible to fit
the last measured data point (Epoch 11 of Barbary et al.).} We note that for this redshift, the lack
of H or He features in the spectrum is an issue (\S 2). We also note that dust formation
may play a role in the decline from peak (\S 3.3).

%s4.3
\subsection{{\it Special case 2: $z=0.57$}}\label{z=0.57}

We now follow the same argument as before for the case of redshift
of 0.57. The results of this fit are shown in the bottom panel of Figure 2.  
The best fit gives an estimate
of $M_{Ni}=4.83~M_{\odot}$ and
$M_{ej}$=19.68 ($\frac{\kappa}{0.1~ \rm cm^{-2} \rm g^{-1}})^{-1}$
($\frac{v_{ph}}{10,000~\rm km~ \rm s^{-1}})$~M$_{\odot}$.  
Note that the ejecta mass is less in this case than for z=0.143
{\bf because the dilation reduces the width of the light curve in the rest frame.}
Again the ejecta mass is rather large,
%JCW addition
but scales with the uncertain opacity and velocity.
The nickel mass in this case is higher, as expected, but within the
range of values predicted for other luminous SNe based on radioactive
diffusion models (Quimby et al. 2007; Smith et al. 2007). The best fitting
gamma-ray leakage parameter in this case yields a gamma ray opacity of
$\kappa_{\gamma}=0.03$ $cm^{2}~g^{-1}$ which is in agreement with the generally assumed
value of 0.03 $cm^{2}~g^{-1}$. Gamma-ray leakage has no substantial effect on the light
curve.

The total ejecta mass in this case implies a kinetic energy equal to
$E_{KE}=1.2 \times 10^{52}$ ($\frac{\kappa}{0.1~ \rm ~cm^{-2} ~\rm g^{-1}})^{-1}$
($\frac{v_{ph}}{10,000~\rm km~ \rm s^{-1}})^{3}$
erg, with a total radiated energy of $E_{rad}=1.4 \times 10^{51}$ erg.
%Within bolometric corrections this value is almost equal to the total kinetic
%energy value as it would be required by the conservation of energy.
%This leads us to conclude that the redshift of z=0.57 is more likely to be the
%real case.
% JCW addition
While the radiated energy is somewhat large for a normal supernova,
the kinetic energy is again quite large, and comparable to that
estimated for z = 0.143. For the fiducial parameters in this case, 
the implication is again that much of the initial shock energy must 
have been lost to adiabatic expansion. 
If the opacity were larger than adopted here for illustration and
the photospheric velocity somewhat smaller, {\bf as implied by the
recombination models}, the estimated kinetic energy
could be more representative of normal supernovae and the estimated 
radiated energy could be comparable to the estimated kinetic energy,
in which case substantial adiabatic losses would not be implied.
This would imply a large initial radius and thus perhaps a dense 
circumstellar medium as will be discussed below.

As for the lower redshift case, the basic radioactive diffusion model does not produce
the steep post-maximum decline. The dashed line in the bottom panel of Figure 2 shows a model with 
recombination at 10,000~K {\bf and $\kappa=$0.2~$cm^{2}~g^{-1}$ which is appropriate for a 
H-poor envelope (see Table 2).} Dust formation might also play a role (\S 3.3).
We expect no evidence for H to be observed at this redshift (\S 2).

%s3.0
% SECTION 5
\section{CSM INTERACTION}\label{csmssd}

Now we investigate the possibility of the contribution of CSM
interaction producing optical/NIR emission in the observed light curve
of SCP06F6.
%JCW addition
There are two versions of this circumstance.  In the model of
Smith \& McCray (2007), the CSM shell is optically thick
and the light curve is controlled by diffusion. Alternatively,
the CSM could be optically thin enough to radiate the
shock collision energy ``instantaneously," but optically thick
enough to convert the shock energy to optical radiation.
%In order for CSM interaction to
%be efficient, a highly opaque CSM environment is required so that
%the shock fronts generated by the collision of the ejecta with the
%CSM increase the density and the temperature sufficiently to
%diffusively radiate a significant amount of energy.

%JCW addition
In the case where the energy release is dominated by diffusion,
the energy to power the light curve arises in the
putative collision of the supernova ejecta with the extended
dense CSM. {\bf There is no need for any radioactive input to power
the light curve near maximum light, although some contribution
from that source cannot be ruled out.} This class of model gives
no simple way to estimate the maximum luminosity in terms of
physical input parameters.  Those input parameters would be
the initial radius and kinetic energy of the underlying supernova
and the initial radius and density distribution of the CSM. An
extreme version of this class of model is obtained with the
assumption that the initial radius of the circumstellar shell
was not far from the observed radius of the photosphere at
maximum light so that adiabatic losses within the shell are
assumed to be minimal. In this case, the post-shock thermal
energy content of the CSM envelope should be roughly comparable
to the radiated energy and to the kinetic energy and both
should be roughly comparable to the initial supernova shock
energy, the kinetic energy of the underlying supernova
(within factors of two). These aspects mean that, except
for this ``extreme" version, there is no simple way to
determine a physical parameter analogous to the initial nickel
mass in the radioactive diffusion model derived from the light curve
peak.  On the other hand, this model still assumes that the luminous
output derives from the diffusion of thermal energy from within
the optically-thick ejecta. In this case, the mass of the
ejecta that determines the width of the light curve is roughly
the sum of the underlying supernova ejecta and the mass of
the shocked CSM matter (Smith \& McCray 2007), and the tools
presented in \S 3 still allow a determination of this
effective mass as a function of redshift. 

If the optical depth is small enough that the diffusion time
for the release of post-shock energy is short, then the
luminosity released as the CSM is shocked is given by 
(Ofek et al. 2007; Gezari et al. 2008; Smith et al. 2008):

\begin{equation}
L \sim 2 \pi \rho_{CSM} R^{2} v_{sh}^{3}
\end{equation}
{\bf where $\rho_{CSM}$ is the density of the CSM and $v_{sh}$ is the velocity
of the shock.}
Note that in this case, the luminosity is presumed to reflect
the local density, and hence the shape of the light curve is
given by the density distribution. The light curve can be
reproduced by a suitable, although entirely {\it ad hoc},
assumption of a density profile. 
%JCW new text
We also note that while some 
information on the mass ejected into the CSM can be estimated, 
this model provides no useful separate information 
on the ejecta mass analogous to the constraint of the rise time to maximum. 
There is thus also no independent constraint on the associated 
kinetic energy.

For the CSM diffusion models, the width of the light curve
yields an estimate of the diffusion time and hence the
ejecta mass from Equation 4. An estimate of the radius of
the shell from kinematics or emissivity (Figure 5)
then yields an estimate of the mean density, ${\bar\rho}$.
For the z=0.143 case the radius at maximum light is estimated
from the kinematics to be $R_{k, 0.143} = v_{ph} t_{max}/(1+z)
= 7.5 \times 10^{15}$ cm for a photospheric velocity of
10,000 km s$^{-1}$. Assuming a temperature of 5000 K in the
rest frame of the supernova (Gaensicke et al., 2008),
the black body radius at maximum light is $R_{bb, 0.143}
= 1.8 \times 10^{15}$ cm. 
%JCW addition
The rather large discrepancy in these two methods for estimating
the radius near maximum light suggests a basic inconsistency
in the diffusion models for this smaller redshift.
Following the same method for
the z=0.57 case, we find that the blackbody radius is
$R_{bb, 0.57} = 6.2 \times 10^{15}$ cm while the kinematic
radius at maximum light is $R_{k, 0.57} = 5.5 \times 10^{15}$ cm.
This represents generally good agreement. If the true blackbody 
radius at maximum light were much higher, this model might be 
strained. 

The significantly low
fraction of radiation energy compared to the kinetic energy for the model
at z = 0.143, $E_{th}/E_{kin} \sim 0.005$ (\S 4.3), implies that there must have been
substantial adiabatic losses for this CSM diffusion model to be viable,
unless the opacity is substantially larger than the fiducial value of
$\kappa = $0.1 cm$^{2}$ g$^{-1}$ (see Figure 4) and the photospheric
velocity substantially lower. 
%Note that for a
%putative hydrogen-rich CSM, the opacity could be substantially higher.
For the case at z = 0.57, the radiated energy is again estimated to be
small compared to the kinetic energy, but rather modest changes in the
fiducial parameters (larger opacity, smaller velocity) could make the
radiated energy comparable to the kinetic energy.  In this case, the 
light curve could be accounted for by the collision of an underlying
supernova with a large, dense, CSM shell. For this situation, with
negligible adiabatic losses, the energy of the underlying supernova 
would be about twice the radiated energy, or about $3\times10^{51}$ ergs.
%If the mean opacity were larger, then the radiated energy would
%be greater than the kinetic energy and the model loses credibility.         
The estimate of {\bf optical depth} near maximum light, $\tau_{max} \simeq 30$, from Equation 5
also applies to this diffusion model. This constant value
of opacity is given as the horizontal dashed line in Figure 6.
Because a CSM envelope is likely to be of
relatively large opacity, the shock diffusion model may be
appropriate to this redshift. The expected composition of the CSM shell is hydrogen
so the lack for H features in the spectrum is a constraint at z=0.143 but not
necessarily at z=0.57 (\S 2).

An estimate of the optical depth can also be made in the context
of the model for which the optical depth is modest and the shock
energy is radiated ``instantaneously." In this case, we can
estimate the density with which the shock collides near maximum
light from Equation 5 as:
\begin{equation}
\rho(R_{max}) \sim \frac{L_{max}}{2\pi R_{max}^{2} v_{sh}^{3}}, 
\end{equation}
%We are thus led to estimate
%the optical depth of the CSM at maximum light which is of the order of
%$\tau \sim \kappa \rho_{max} R_{max}$ with $\rho=mn$, $n$ being
%the number density and $m$ the H+He mass ($2.1 \times 10^{-24} gr$).
%[HOW DO WE KNOW WE ARE DEALING WITH H AND HE?]
%For the number density of ejecta we have $n \sim L/2\pi R^{2} v_{ph}^{3}m$
%assuming a constant density and that the shock energy is thermalized and
%rapidly diffusively emitted,
Taking the shock velocity to be
constant and equal to the photospheric expansion velocity
and hence $R_{max}=v_{ph}t_{max}$ yields:
\begin{equation}
\tau \sim \frac{\kappa L_{max}}{2 \pi v_{ph}^{4} t_{max}},
\end{equation}
with $t_{max} = t_{m,obs}/(1 + z)$. Note that the optical 
depth for this model is very 
sensitive to the assumed shock velocity. Adopting a
characteristic shock velocity of 10,000 km~s$^{-1}$,
we can estimate $\tau$ for a given opacity.
%if we adopt for temperature the value of Gaensicke et al. (2008) (5000K).
In Figure 6 we also show how the optical depth at maximum light
estimated in this way scales with redshift. Note that the optical
depth estimated based on this model increases rapidly with redshift.
%This means the range of validity of this model is limited. 
For optical depths less than unity, the luminosity is expected to be emitted
at high-energy, non-optical wavelengths. For optical depths much
greater that unity, the diffusion time, not shock propagation, will
control the emission timescale. From Figure 6, the range of validity of
this model for the fiducial parameters is $z \geq 0.9$. 
%a redshift less than about 0.1, suggesting that this class of model 
%is not applicable to SCP06F6, if it is some variety of supernova.
%For redshifts greater than $\sim$0.1 for $\kappa = 0.05$ $cm^{2}~g^{-1}$
%the CSM becomes optically thick and CSM interaction can account
%for the lightcurve of SCP06F6. 
%Figure 7 gives $\tau \sim 0.56$
%for $z=0.143$ and $\tau \sim 12.33$ for $z=0.57$.
%, so the constraints on the model
%may be even tighter than we have indicated. The optical depths would
%be less if the shock velocity were substantially larger than $v_{ph}$.
%
%These results indicate that the CSM might be marginally optically
%thick for the z = 0.143 case, if the opacity were modest. If
%the opacity were higher than $\kappa \sim 0.1$ $cm^{2}~g^{-1}$
%as might be appropriate for a hydrogen plasma, the CSM would be
%optically thick and the model would be invalid.
%which means that a pure CSM interaction model is not sufficient
%to contribute to the observed light curve. [BUT WE ARE SOMEHOW
%NEGLECTING THE OPTICAL DEPTH OF THE SUPERNOVA] 
For the z = 0.57 case, this model could be made self-consistent for
higher opacity or smaller shock velocities. 
The lack of evidence for H in the spectrum again constrains
this class of models, especially at lower redshifts.
%the optical depth is large for any reasonable
%estimate of the mean opacity, so the model is not valid. Only diffusion
%models, if any, are pertinent to SCP06F6 at this redshift.
%%of 12.33 that we estimated suggests that the CSM
%environment of SCP06F6 could be rather opaque. A shocked CSM might
%play a role in this case.

We can also constrain the mass loss rate associated with the CSM.
For circumstances in which the CSM is optically-thick and
diffusion controls the light curve, we can use the estimates
of the CSM mass and the radius of the configuration to make
an estimate of the effective mass loss rate. The mass loss
rate can be estimated as:
\begin{equation}
\dot{M}=\frac{M_{csm}}{t_{max}v_{ph}/v_{w}},
\end{equation}
where $v_w$ is the velocity of the ``wind" that led to
the formation of the CSM. 
For a red-giant type wind, this
velocity might be 10 km s$^{-1}$. For an LBV-type mass loss
event, a typical velocity might be $\sim$ 100 km s$^{-1}$ (Smith et al.
2004; Smith 2006).
If we assume that the diffusion time is dominated by the mass
ejected into the CSM prior to explosion, and hence that
the ``ejected" mass in Figure 2 is a measure of the CSM mass,
then we can estimate the effective mass loss rate.
For z = 0.143, we get 0.15~M$_{\odot}$~ yr$^{-1}$ and
1.5~M$_{\odot}$~yr$^{-1}$ for $v_{w}$ equal to 10 and 100 km s$^{-1}$,
respectively. For z = 0.57, we get 0.11~M$_{\odot}$ yr$^{-1}$ and 
1.11~M$_{\odot}
$~yr$^{-1}$. These values were all based on our fiducial opacity
and so scale as $(\frac{\kappa}{0.1~ \rm cm^{-2}~\rm g^{-1}})^{-1}$.
For a Wolf-Rayet progenitor the wind velocity could be up
to 1000 km s$^{-1}$ implying an even larger mass loss rate.
All these values are large, implying a more LBV-like process
if this diffusion model is pertinent.

%JCW adapted old text
We can also estimate the mass loss rate implied in the model
where the luminosity arises by shock interaction in a 
CSM of modest optical depth. The total envelope mass within a 
radius $R = v_{sh} t_{max}$ is, with Equation 7: 
\begin{equation}
M_{csm} = \frac{4}{3} \pi R^{3} \rho \simeq \frac{2}{3} \frac{L_{max}
t_{max}}{v_{sh}^{2}}.
\end{equation}
With Equation 9, we can then write:
\begin{equation}
\dot{M}= \frac{2}{3} \frac{L_{max} v_{w}}{v_{sh}^{3}} 
   \sim 10^{-4}~ M_{\odot} ~ yr^{-1} \frac{L_{max, 42} v_{w,100}} {v_{sh,10,000}^{3}},   
\end{equation}
where $L_{max, 42}$ is the peak luminosity in $10^{42}$~erg s$^{-1}$, $v_{w,100}$ is the
wind velocity in units of 100 km s$^{-1}$ and $v_{sh,10,000}$ is the shock
speed in units of 10,000 km s$^{-1}$.  
For a redshift of 0.57, the corresponding mass loss rate is about 
$3\times10^{-3} v_{w,100}~v_{sh,10,000}^{-3}$ M$_{\odot}$ yr$^{-1}$. 
This mass loss rate could be representative of a Wolf-Rayet star
progenitor, but then the wind velocity should be higher to correspond
to observed Wolf-Rayet stars. 
%JCW text
Another difficulty is that this mass distribution cannot be a standard 
$\rho \propto r^{-2}$ wind as roughly expected for a Wolf-Rayet star. 
Rather, the density profile must be carefully ``designed" to reproduce 
the shape of the light curve. This conceptual problem mitigates against 
this model on general grounds. 

Gaensicke et al. (2008) reported a very high 0.5 - 10 keV X-ray flux
near optical maximum of about 10$^{-13}$~erg~s$^{-1}$~cm$^{-2}$.
This is nearly four times the bolometric flux. It is possible that
the reported detection should instead be treated as an upper limit
(D. Pooley, private communication). 
The X-ray flux could arise from shocked optically-thin
CSM. In this case Equations 5 and 7 pertain, and if we take the reported
X-ray flux at face value, we estimate
a particle density $6.4 \times 10^{6}$ cm$^{-3}$ and 
$2.6 \times 10^{8}$ cm$^{-3}$ for redshifts of
0.143 and 0.57, respectively, for a shock velocity of 10,000~km~s$^{-1}$.
This density is 100-10,000 times smaller than the average density ${\bar\rho}$ 
of a dense optically-thick shell for the two redshift cases.
This result implies that the observed X-ray flux, if real, might possibly 
arise in the outer, lower density portion of the diffusive shell that 
produces the optical display. For a CSM of modest optical depth, it 
is difficult to see how a self-consistent model based on shock interaction
could simultaneously explain the optical and the X-ray luminosity.
A possibility is that the medium is clumpy with dense clumps providing
the optical emission and inter-clump regions providing X-rays.

\section{DISCUSSION AND SUMMARY}\label{disc}

We discussed the applicability of some widely used SN light
curve models in reproducing the observed light curve of the
luminous peculiar transient SCP06F6 discovered by Barbary et al.
(2008). The parameters estimated based on various models
are summarized in Table 1. The observed light curve can be approximated
by a smooth simple diffusion model, but the rapid decline from maximum is not
consistent with the simple models.
%JCW addition
Fits to a radioactive decay diffusion model provide
estimates for the ejecta mass, the nickel mass, and the gamma ray opacity versus redshift
depending on key, but uncertain parameters, especially the mean
optical opacity {\bf and photospheric velocity.} We also note that while the data invite the interpretation of a smooth
rise and decline of the light curve, the photometric data are sparse. The possibility
of a precursor peak, such as displayed by SN1987A or SN2005bf, cannot be ruled out. The
existence of such a feature would modify any interpretation of the data.

We considered two specific choices of the redshift,
$z =$ 0.143 based on a suggestion by Gaensicke et al. (2008)
and $z =$ 0.57, based on a crude fit to Ca H\&K and iron peak
absorptions in various supernovae. Substantially higher redshifts,
greater than $\sim 1.1$,
do not lead to self-consisent results in the context of the
radioactive decay diffusion model: the nickel mass becomes too
large compared to the ejecta mass and the radiated energy exceeds
the kinetic energy. For these two specific choices of redshift, we
found nickel masses of 0.28 and 4.83~M$_{\odot}$, respectively,
and ejecta masses of $\sim$37 and $\sim$20 M$_{\odot}$, respectively, 
for an adopted mean opacity of 0.1 $cm^{2}~g^{-1}$ and a photospheric
velocity of 10,000 km s$^{-1}$. The estimated ejecta mass scales 
inversely with the opacity and directly with the velocity. The results
are roughly commensurate with normal core-collapse supernovae
at lower redshift, but at the larger redshift, the nickel
mass would require a different situation,
perhaps similar to that invoked for some models of SN~2006gy   
based on pair-formation supernovae (Smith et al. 2007).
For both redshifts, the implied kinetic energy is very
high, $\sim 10^{52}$ erg s$^{-1}$ for the fidicial parameters, {\bf but would be less for smaller
ejecta velocities.} 
In either case, our simple diffusion models do not acount for the rapid post maximum
decline. Some other factor would need to be invoked such as a change in opacity or dust formation.
We show that recombination of H, maybe He, might plausibly account for a decline in opacity and hence in the
light curve.
%Table 1 summarizes the estimated parameters of our models for the two
%redshift cases that we investigate.
%This suggests that if SCP06F6 is a supernova, it may have
%some extreme properties.

We also consider models in which the optical luminosity is
provided by collision of the supernova with a dense CSM
shell in the spirit of Smith \& McCray (2008) or models
in which the shock energy is thermalized, but rapidly
radiated away (Ofek et al. 2007; Gezari et al. 2008;
Smith et al. 2008). In general, the
models with dense, optically thick extended shells behave
in a manner similar to that of the radioactive decay diffusion
models, except that the energy to heat the CSM envelope is
presumed to derive from the kinetic energy of the underlying
explosion, rather than radioactive decay. These models applied
at low redshift give the radiated energy substantially
less than the kinetic energy, implying that to be self-consistent,
the CSM envelope must have undergone substantial adiabatic
expansion after being shocked. The same is true for the model
applied at larger redshifts, but reasonable choices of the
parameters would yield a model for which the shell had
expanded little by maximum light. For such choice of parameters,
the energy of the underlying explosion could be more modest. 
%The models are moderately successful at the upper
%limits to redshift, z $\sim 0.5$ beyond which they demand
%a radiated energy exceeding the kinetic energy and hence
%are invalid. 
These models demand substantial mass in the CSM
and hence effective mass loss rates that are reminiscent of
LBV mass-loss episodes. Models in which the CSM is only modestly
optically thick so the energy can be thermalized but also
radiated rapidly could be applicable at higher redshifts. Such models might
be consistent with mass loss rates reminiscent of Wolf-Rayet stars,
but to match the shape of the light curve, the density profile
would need to be carefully contrived, an unlikely happenstance. 
%They might be made
%modestly self-consistent at low redshifts and for low
%mean opacities, but do not meet the optical depth
%constraints for higher redshift or higher opacity. A somewhat
%lower density CSM might account for the observed X-ray
%flux, but the constraints suggest that even then matter is dense and perhaps
%associated with the thinner outer portions of a CSM associated
%with an episode of heavy, LBV-like mass loss.

%For the spectral choice for the redshift, the values for the ejecta
%mass in both of the cases exceed 8 solar masses, which means that
%the progenitor of the explosion was a massive Red giant or Wolf Rayet star.
%The nickel mass estimates range from 0.015 to 0.38 solar masses, a
%range characteristic for other Type II supernovae explosions.

The spectrum of SCP06F6 does not resemble any other transient spectrum
that has been obtained so far. The four broad absorption features can
be reasonably well fit by the molecular C$_2$ Swan bands of a carbon-rich
star of temperature 3000-10,000K (Gaensicke et al. 2008).
%JCW addition
Swan bands are also observed in white dwarfs at substantially higher
temperatures, but only because of the very high gravity that
would not pertain to supernova ejecta. Pending the calculation of
a realistic supernova atmosphere model showing that C$_2$ can form and survive,
we find this hypothesis intriguing, but unlikely, at least around maximum light
when the ejecta is supposted to have $T_{eff} \simeq$~5000K. On the other
hand, at later phases, when the ejecta has cooled, the formation of C$_{2}$ and other
molecules, even dust grains, is indeed a possibility {\bf although this would imply
the presence of CN ($\lambda$7900 \AA\ rest frame) which we do not detect in the co-added
VLT, Subaru and Keck spectra of SCP06F6}.
%In addition, the
%high X-ray flux may also be expected to destroy Carbon molecules.

%In the absence of late-time photometry of the transient we cannot
%obtain a more accurate measure of the initial nickel mass that is
%expected from $^{56}$Co radioactive decay (using Sutherland \& Wheeler 1984)
%in a putative radioactive tail that should have been present given the implied
%large ejecta mass and efficient gamma-ray trapping.

%Within the radioactive decay diffusion type of models, we explore the
%scaling of the the set of parameters ($M_{Ni}$,$M_{ej}$,$E_{th}$,$E_{kin}$,
%and $\tau_{max}$) with redshift which puts constrains on the actual
%redshift of the object. This analysis shows that values of redshift
%above $\sim$0.68 are excluded as implausible, since above this redshift
%the energy is not conserved and the nickel mass becomes unrealistically
%high and exceeds the ejecta mass. For redshifts above 0.2 the CSM at
%maximum light is optically thick and a shocked CSM or ejecta-CSM interaction
%model can account for the observed light. This analysis shows that a
%redshift of 0.143 that was proposed by Gaensicke et al. (2008) SCP06F6
%has an optically thin CSM envelope. The lightcurve of SCP06F6 at late
%times though declines slower than Co-decay (see figures 5 and 6) and
%this means that there is an additional source of energy that powers
%SCP06F6 at these times. 

%JCW addition
We propose that a model of SCP06F6 worth further consideration 
is one in which the event is at a redshift of z $\sim$ 0.57 for
which the blue absorption features are Ca H\&K and iron-peak
absorption features as qualitatively seen in the UV of some
nearby supernovae. 
%at late times CSM interaction can play a role if we accept that
%the redshift of the transient is 0.57.  At this redshift we are
%able to identify two of the three broad absorption features of
%the spectrum of SCP06F6 by comparison with SN1992A as blended iron-peak
%elements.
At this redshift the light curve can be reproduced semi-quantitatively
with a diffusion model based either on radioactive decay requiring
several solar masses of nickel or collision of a moderately
energetic supernova with a dense CSM envelope.
%is also consistent with our energetic arguments (Figure 2) and
%The detection of high X-ray flux reported by Gaensicke et al. (2008)
%close to maximum light might be explained by the collision of the
%outward propagating shock with the outer portions of the CSM structure.
%We propose that this high X-ray flux can be produced through a specific
%CSM-ejecta shocked interaction. The implied high mass-loss rate at
%$z$=0.57 leads to the conclusion that the progenitor of SCP06F6 was
%probably a Wolf Rayet type of star with strong winds before the explosion.
%We note that the X-ray behavior of a number of proposed ``ultrabright"
%supernovae has been rather varied (that from SN2006gy was
%rather low), suggesting that the X-ray properties
%are a particularly interesting constraint on the underlying events.

%Our results indicate that SCP06F6 has the characteristics of
%a typical Type II supernova originated from a Wolf Rayet type
%progenitor at $z$=0.57 which shed a significant amount of ejecta.
%The high X-ray flux can be generated if we accept a high mass-loss
%rate and a particular configuration of shocked CSM-ejecta interaction.
The characteristics of SCP06F6 that are deduced from its light curve
at $z$=0.57 indicate that its luminosity is one order of magnitude
less than that of the exceptionally luminous 2005ap, 2006gy, 2008es
and 2005tf. This means that if SCP06F6 is a supernova fitting 
the characteristics presented here, defining a new class of objects
is not necessary to account for it. It fits well within the
rubric of these other bright events.

% SECTION 6

We are grateful to Kyle Barbary, Robert Quimby, Andy Howell, {\bf Nick Suntzeff}, 
Alicia Soderberg, Dave Pooley,
Milos Milosavljevic and Volker Bromm for useful comments and suggestions {\bf and to the anonymous 
referee whose insightful queries substantially improved the manuscript.}
This research is supported in part by NSF Grant AST-0707669 and by
the Texas Advanced Research Program grant ASTRO-ARP-0094. 
E. Chatzopoulos would like to thank Propondis foundation
of Piraeus, Greece for its support of his studies. J. Vinko received support
from the Hungarian OTKA Grant K76816.
%%%%%%%%%%%%%%%%%%%%%%%%%%%%%%%%%%%%%%%%%%%%%%%%%%%%%%%%%%%%%%%%%%%%%%%%%%%%
%%            REFERENCES
%%%%%%%%%%%%%%%%%%%%%%%%%%%%%%%%%%%%%%%%%%%%%%%%%%%%%%%%%%%%%%%%%%%%%%%%%%%%

%\begin {references}

{}                           
%\end{references}

%%%%%%%%%%%%%%%%%%%%%%%%%%%%%%%%%%%%%%%%%%%%%%%%%%%%%%%%%%%%%%%%%%%%%%%%%%%%
%%            FIGURES 

\begin{figure}
\begin{center}
\includegraphics[angle=270,width=18cm]{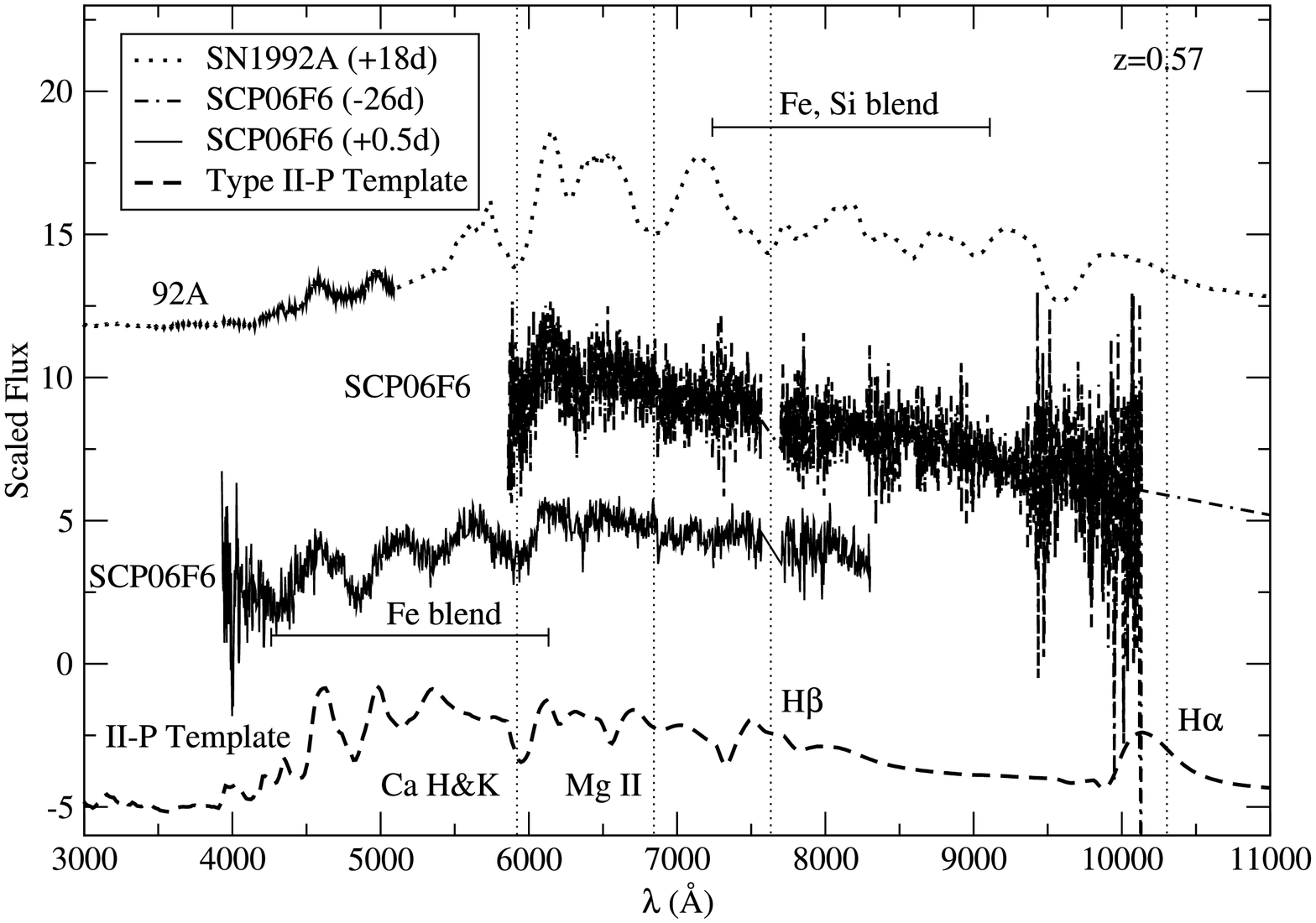}
\caption{ Comparison of the VLT and the Subaru observer's frame spectrum of SCP 06F6 obtained
on 05-18-2006 (near maximum) and  05-22-2006 respectively (Barbary et al. 2008) 
with the IUE+CTIO spectrum of the Type Ia SN~1992A obtained on 01-24-1992 (+18d after maximum)
(Kirshner et al. 1993) and with a template Type-IIP spectrum at +6d after
the explosion (Gilliland et al. 1999) both 
boosted to a redshift of 0.57. The dashed vertical
lines indicate the positions of the Ca H\&K absorption component and the
Mg II triplet. The ranges of Fe and Si blends are also indicated. Line
identification is based on Kirshner et al. (1993). The redshifted positions
of the H$\alpha$ and H$\beta$ lines are also indicated for illustration. Note
that H$\alpha$ and H$\beta$ could be present, but unobserved, at this redshift
due to the large redshift and telluric contamination.}
\end{center}
\end{figure}

\begin{figure}
\begin{center}
\includegraphics[angle=270,width=18cm]{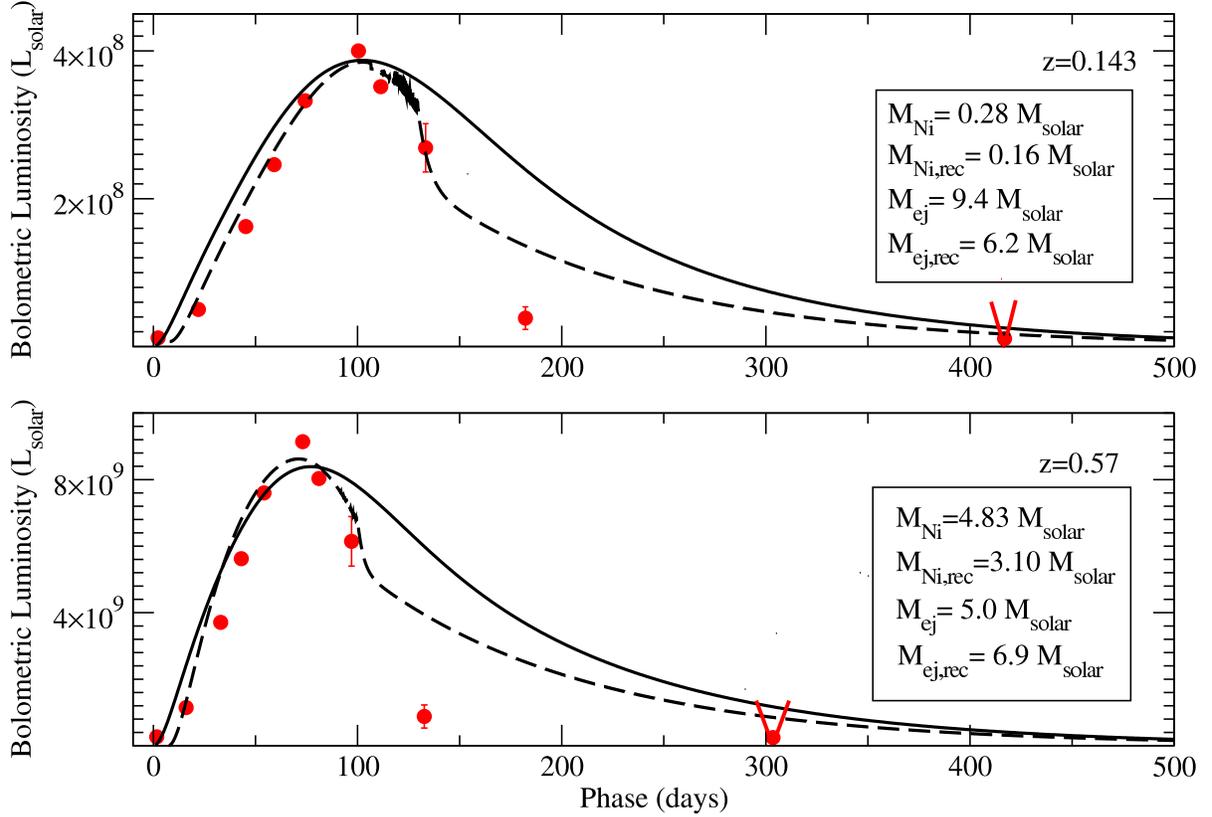}
\caption{The rest frame quasi-bolometric luminosity light curve of SCP 06F6 (\S 3.1) at
redshifts of 0.143 (top panel) and 0.57 (bottom panel) (solid points). In each case,
the solid line is a simple radioactive diffusion model (\S 3.1) and the dashed line an illustration of 
the Arnett \& Fu (1989) model that includes the effects of recombination giving
a decline in the optical opacity and thus a rapid post-maximum decline.
The derived values of the original nickel mass and of the ejecta mass are given
in the insets. {\bf Note that in this figure the basic model assumes 
$\kappa =$ 0.1 cm$^2$ g$^{-1}$ and the corresponding recombination models assume 
$\kappa =$ 0.4 and 0.2 cm$^2$ g$^{-1}$, respectively, for z = 0.143 and z = 0.57 (see Tables 1, 2).}}
\end{center}
\end{figure}

\begin{figure}
\begin{center}
\includegraphics[angle=270,width=18cm]{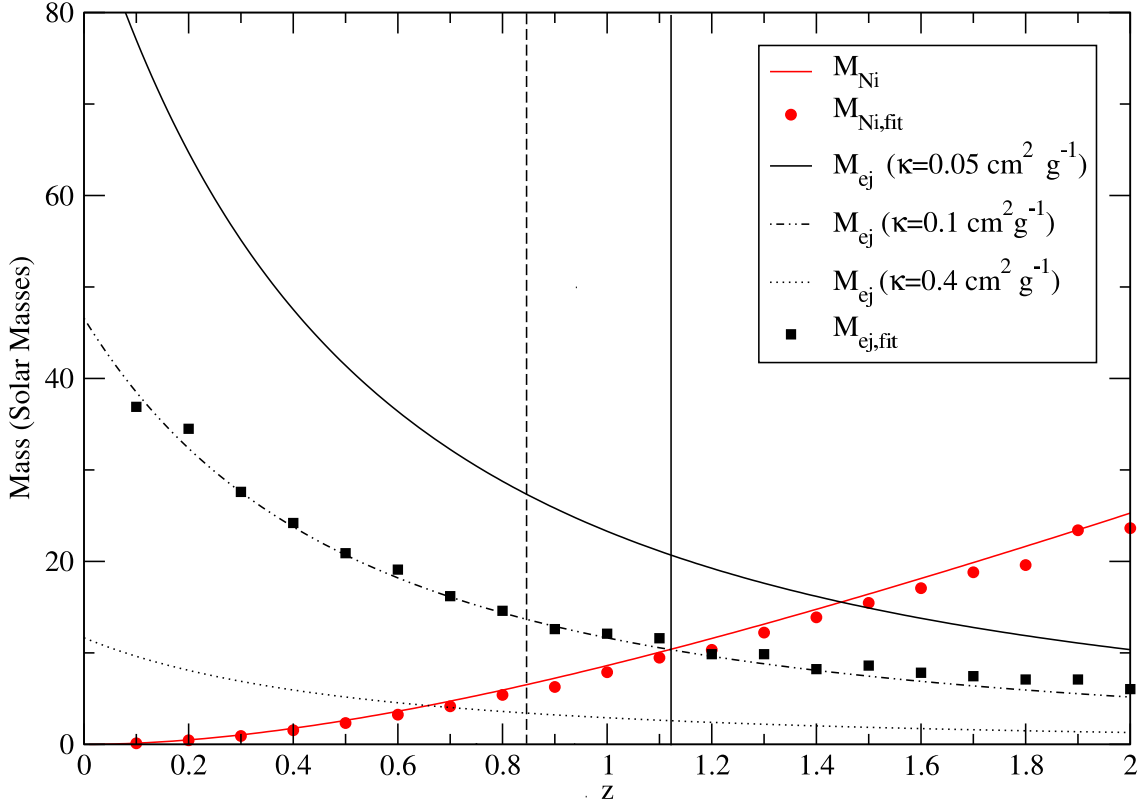}
\caption{The dependence of the initial nickel mass (rising line) and
the ejecta mass (decreasing lines) on redshift for SCP06F6. The line for the nickel mass
is based on Equations 1 and 2. The lines for the ejecta mass are from Equation 4. The
ejecta mass depends linearly with the photospheric velocity assumed 
to be 10,000 km s$^{-1}$ and inversely on the mean opacity, for which
three choices are shown: $\kappa = 0.05 $ (solid) $, 0.1 $ (dotted dashed) $, 
0.4 $ (dotted) $ $ cm$^{2}$ g$^{-1}$. The filled circles and squares correspond to the
best fitting estimates of the nickel mass and the ejecta mass respectively calculated
by the Monte-Carlo $\chi^{2}$ minimization code (see text).
At redshifts to the right of the solid vertical line the mass
of radioactive nickel becomes larger than the total ejecta mass
for $\kappa = 0.1$ cm$^{2}$ g$^{-1}$. This line thus defines the lower
redshift boundary of a ``forbidden region" in redshift space
for this class of radioactive diffusion model. The dashed
vertical line indicates the redshift at which the nickel mass
is half the total ejected mass for $\kappa = 0.1$ cm$^{2}$ g$^{-1}$. 
The forbidden region extends to lower redshift for higher opacity 
and lower photospheric velocity.}
\end{center}
\end{figure}

\begin{figure}
\begin{center}
\includegraphics[angle=270,width=18cm]{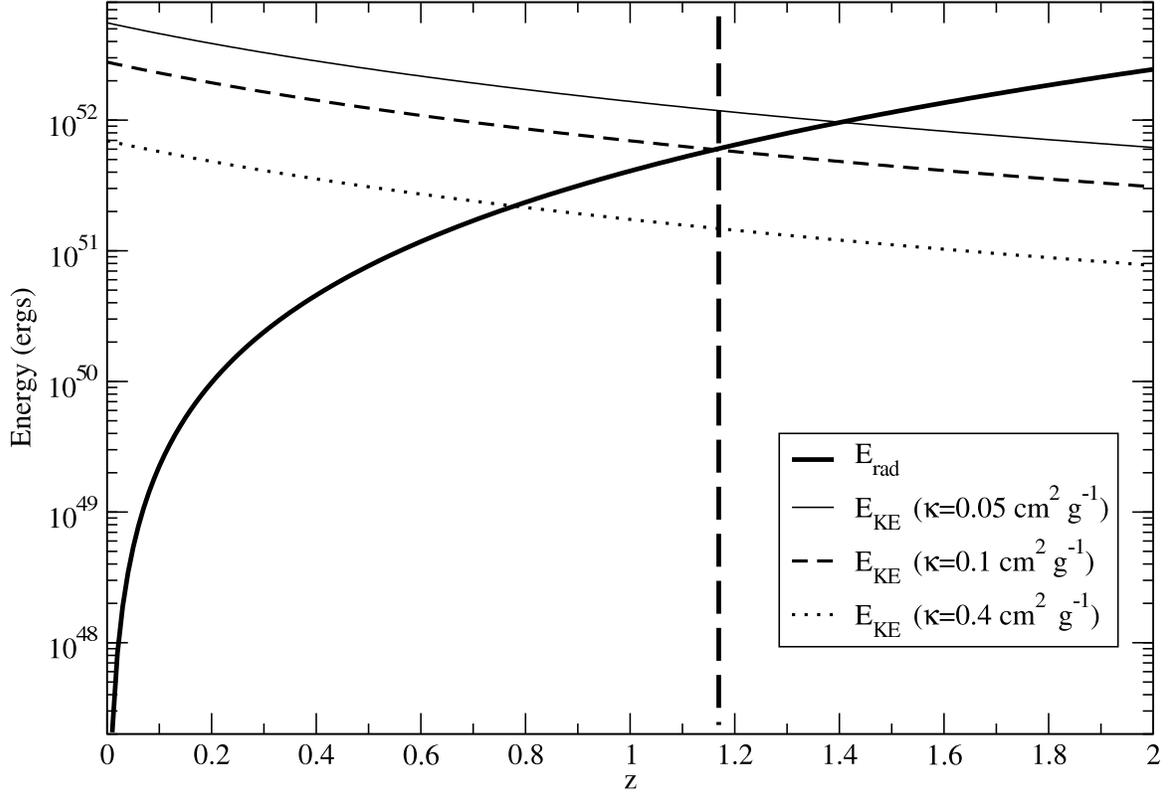}
\caption{The dependence of the total radiated energy (thick solid line) and the kinetic
energy of the ejecta on redshift for SCP06F6.
Three choices of the mean opacity are shown for the
kinetic energy (see Figure 1). The region at redshifts to the
right of the dashed vertical line at $z=1.18$ is forbidden 
for $\kappa = 0.1$ cm$^{2}$ g$^{-1}$ since
the radiated energy becomes larger than the kinetic energy of the ejecta.
This forbidden region extends to lower redshifts for larger opacity.}
\end{center}
\end{figure}

\begin{figure}
\begin{center}
\includegraphics[angle=270,width=18cm]{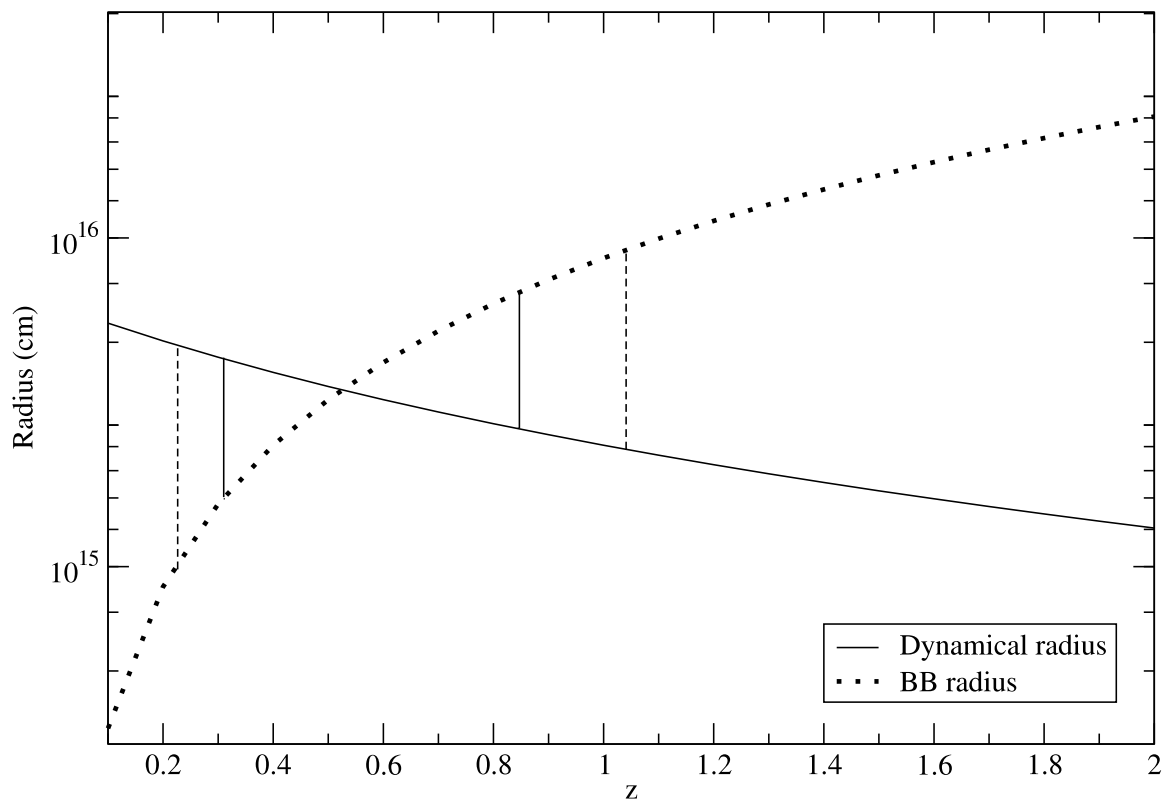}
\caption{The dependence on redshift of the photospheric radii
based on kinematics (R = v$_{ph}$ t) and on black body emission  at maximum
light. 
The kinematic radius
assumes a constant photospheric velocity in the rest frame
of 10,000 km s$^{-1}$. The black body radius assumes a
constant temperature of 5000 K in the rest frame. The
vertical lines indicate the redshift range over which these
estimates of the radius agree within a factor of two (solid line) and a
factor of three (dashed line), respectively.}
\end{center}
\end{figure}

\begin{figure}
\begin{center}
\includegraphics[angle=270,width=18cm]{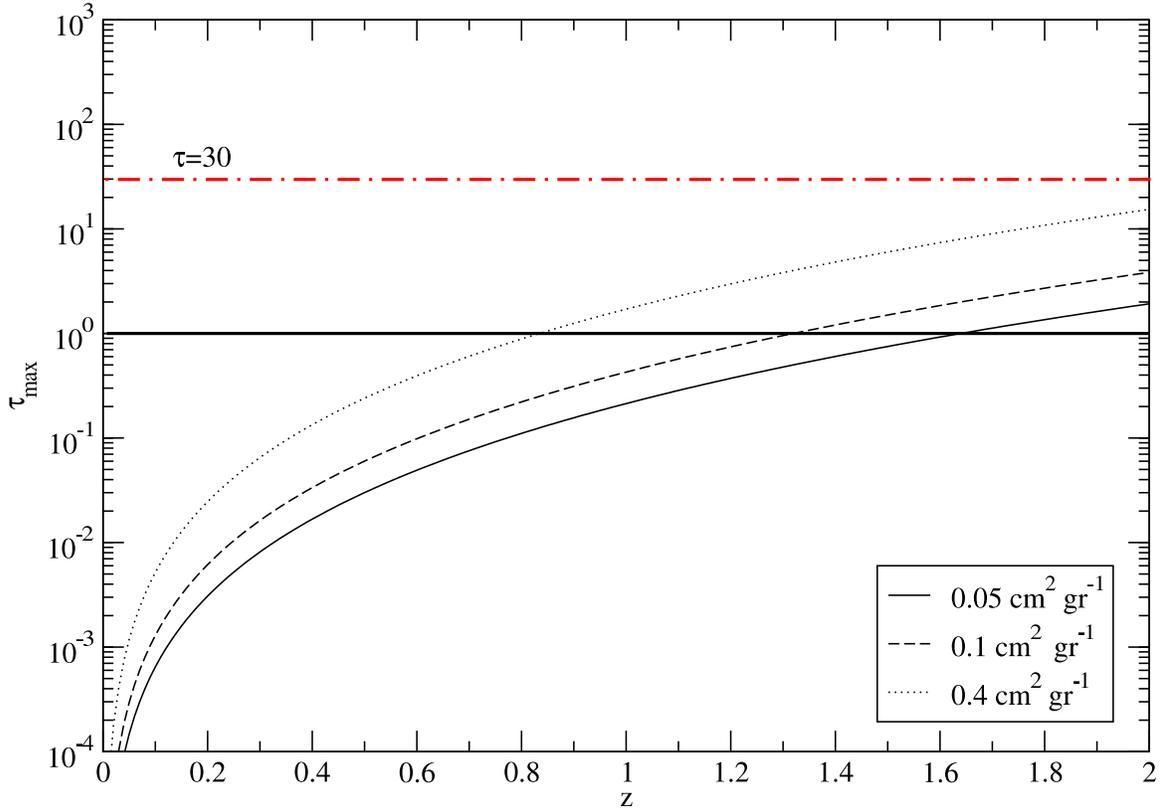}
\caption{The dependence of the optical depth at maximum light on redshift
for CSM models of SCPO6F6. The horizontal dot-dash line at $\tau = 30$ 
represents
the estimated optical depth ($\tau$ = c/v$_{ph}$) for a diffusion
shell model with radius given by the kinematic estimate (see text). 
The solid horizontal line corresponds to $\tau=1$. The
thin solid, dashed and dotted lines are based on a model of moderate 
optical depth for which the luminosity is quickly radiated behind the 
shock. This model is valid only for $\tau \gtsim 1$.}
\end{center}
\end{figure}

%%%%%%%%%%%%%%TABLE 1%%%%%%%%%%%%%%%%%%%%%%%%%%%%%%%%%%%%%%%%%%%%%%%%%%%%%%%
\clearpage
%\begin{landscape}
\setcounter{table}{0}
\begin{deluxetable}{lllllllllllcccccc}
\tabletypesize{\tiny}
\tablewidth{0pt}
\tablecaption{Summary of the fiducial physical parameters of SCP06F6}
\tablehead{&&\colhead {z = 0.143}&&&\colhead {z = 0.57}&&\\
\colhead {$\kappa/cm^{2}g^{-1}$} 
&
\colhead {0.05} &
\colhead {0.1} &
\colhead {0.4} &
\colhead {0.05} &
\colhead {0.1} &
\colhead {0.4} &}

\startdata
$t_{m,rf}/days$ 	     &  90   &  90  & 90    & 65  & 65 &  65 \\
$M_{Ni}/M_{\odot}$		&  0.28   &  0.28  & 0.28   & 4.83  & 4.83 & 4.83 \\
\hline
$M_{ej}/M_{\odot}$		&  75.4   &  37.7  & 9.4    & 39.4  & 19.7 &  4.9 \\
$E_{th}/10^{50}ergs$	   &  0.91   &  0.91  & 0.91   & 14.0  & 14.0 & 14.0 \\
$E_{kin}/10^{51}ergs$	     &  42.0   &  21.0  & 5.3	 & 5.5   & 11.0 &  2.8 \\
$R_{bb}/10^{15}cm$	    &  1.7   &   1.7  & 1.7    & 7.8   &  7.8 &  7.8 \\
$R_{kin}/10^{15}cm$	    &	7.5   &   7.5  & 7.5	& 5.5	&  5.5 &  5.5 \\
$\tau_{thick}\dagger$	    &  30   &	30  & 30    & 30   &  30 & 30 \\
$\tau_{thin}\dagger$	    &  0.001   &   0.003  & 0.011    & 0.04   &  0.08 & 0.32 \\
$\dot{M}_{thick}/M_{\odot}~yr^{-1}\dagger$&    3.0   &   1.5  & 0.38	& 2.22  &  1.11 & 0.28 \\
$\dot{M}_{thin}/M_{\odot}~yr^{-1}\dagger$&     $2\times 10^{-4}$   &   $2\times 10^{-4}$  & $2\times 10^{-4}$ 
  & $3\times 10^{-3}$  &  $3\times 10^{-3}$ & $3\times 10^{-3}$ \\
\enddata 
\tablecomments{($v_{ph}=10,000~km s^{-1}$ for all cases).
%$\ast$The first three columns correspond to the z=0.143 case
%and the last three to the z=0.57 case.
$\dagger$The values of $\tau_{thick}$ and $\dot{M}_{thick}$ are estimated based on a model of an 
optically thick shell and the values of $\tau_{thin}$ and $\dot{M}_{thin}$ are estimated based 
on a model of moderate optical depth for which the luminosity is quickly radiated behind the shock.
The given mass loss rates assume a ``wind" velocity of 100 km~s$^{-1}$.}
\end{deluxetable}

%\begin{landscape}
\setcounter{table}{1}
\begin{deluxetable}{lccc}
\tabletypesize{\tiny}
\tablewidth{0pt}
\tablecaption{Summary of the parameters of the recombination models}
\tablehead{\colhead {Parameter}&\colhead {z = 0.143}&\colhead {z = 0.57} }
\startdata
%z$\ast$&0.143&0.143&0.143&0.57&0.57&0.57\\
%\hline
$R_{0}/10^{11}cm$              &  2     &  2    \\
$M_{Ni}/M_{\odot}$             &  0.16   &  3.1   \\
$M_{ej}/M_{\odot}$             &  6.2   &  6.9   \\
$E_{th}/10^{51}ergs$           &  1.5   &  1.0   \\
$E_{kin}/10^{51}ergs$          &  1.5   &  3.0   \\
$T_{ion}/K$                    &   5500   &   10000   \\
$\kappa~/cm^{2}g^{-1}$         &   0.4   &   0.2   \\
$v_{mean}/kms^{-1}$                &  3,900   &  2,700 \\
\enddata 
\end{deluxetable}
%\end{landscape}

%%%%%%%%%%%%%%%%%%%%%%%%%%%%%%%%%%%%%%%%%%%%%%%%%%%%%%%%%%%%%%%%%%%%%%%%%%%%%%%

\end{document}